\documentclass[10pt, conference, compsocconf]{IEEEtran}
\usepackage{cite}
\ifCLASSINFOpdf
   \usepackage[pdftex]{graphicx}
\else
\fi
\usepackage[cmex10]{amsmath}
\usepackage[tight,footnotesize]{subfigure}

\usepackage{moreverb}
 \usepackage{booktabs}
\usepackage{graphicx}

\usepackage{fancyvrb}

\usepackage{color,soul}

\usepackage[table]{xcolor}

\usepackage{tabularx}
 
\def \etal {\textit{et al. }} 
\title{Context-based Pseudonym Changing Scheme for Vehicular Adhoc 
Networks}

\begin{document}



\author{
	\IEEEauthorblockN{Karim 
Emara\IEEEauthorrefmark{1}\IEEEauthorrefmark{2}, Wolfgang 
Woerndl\IEEEauthorrefmark{1} and Johann Schlichter\IEEEauthorrefmark{1}}

	\IEEEauthorblockA{\IEEEauthorrefmark{1}Department of Informatics, 		
	Technical University of Munich (TUM), Garching, Germany \\
		Email: \{emara, woerndl, schlichter\}@in.tum.de}
	\IEEEauthorblockA{\IEEEauthorrefmark{2}Faculty of Computer and Information 
		Sciences, Ain Shams University, Cairo, Egypt\\
		Email: karim.emara@cis.asu.edu.eg}

}

%\IEEEspecialpapernotice{This an extended version of the paper \cite{Emara15c}}

\maketitle

\begin{abstract}
Vehicular adhoc networks allow vehicles to share their information for safety 
and traffic efficiency. However, sharing information may threaten the driver 
privacy because it includes spatiotemporal information and is broadcast 
publicly and periodically. In this paper\footnote{This paper is an extended 
version of a previous publication \cite{Emara15c}.}, we propose a 
context-adaptive 
pseudonym changing scheme which lets a vehicle decide autonomously when to 
change its pseudonym and how long it should remain silent to ensure 
unlinkability. This scheme adapts dynamically based on the density of the 
surrounding traffic and the user privacy preferences. We employ a multi-target 
tracking algorithm to measure privacy in terms of traceability in realistic 
vehicle traces. We use Monte Carlo analysis to estimate the quality of service 
(QoS) of a forward collision warning application when vehicles apply this 
scheme. According to the experimental results, the proposed scheme provides a 
better compromise between traceability and QoS than a random silent period 
scheme.

\end{abstract}


% \fi
\IEEEpeerreviewmaketitle

\section{Introduction}
Vehicular adhoc networks (VANET) are those networks formed among vehicles and 
roadside units (RSUs) to provide diverse traffic-related and infotainment 
applications. VANET are envisioned to enhance traffic safety and efficiency by 
increasing the awareness of vehicles about their surrounding traffic. To attain 
this awareness in real-time, vehicles are required to broadcast periodically 
their current state (position, speed, heading, etc.) in authenticated 
\textit{beacon} messages. These messages may threaten the driver location 
privacy when they are collected by an external eavesdropper because vehicle 
trajectories can be reconstructed from these messages. Subsequent beacon 
messages are linkable whether through matching similar identifiers (i.e., 
pseudonyms) and tracking the subsequent spatiotemporal information 
\cite{Emara13a, Emara13w, Wiedersheim10}. Although the exchanged beacons 
contain no identifying information, a de-anonymization of the reconstructed 
anonymous traces is achievable using work/home pairs \cite{Golle09} or top N 
locations \cite{Zang11} and with the help of geosocial networks \cite{Cecaj14}.

There are many pseudonym changing schemes (i.e., privacy schemes) proposed in 
literature that address this linkability issue. The main stream of these 
schemes suggests to preload vehicles with a pool of pseudonyms where a single 
pseudonym is used at a time and changed periodically \cite{Petit14}. However, 
in order to be effective, a vehicle must change its pseudonym simultaneously 
with other nearby vehicles since a sole change within an area can be easily 
linked to the old pseudonym. Although simultaneous pseudonym changes are 
required, they are not sufficient to guarantee  unlinkability. An adversary can 
utilize the spatiotemporal information to relink messages of new and old 
pseudonyms originating from the same vehicle using multi-target tracking 
techniques \cite{Emara13w, Wiedersheim10}. 

Therefore, it is required to change pseudonyms in an unobserved zone in which 
the adversary cannot monitor the vehicle movements. This zone is often realized 
by a silent period \cite{Sampigethaya05} or in a predetermined location (i.e., 
mix-zone) \cite{Freudiger07}. On the one hand, the silent period scheme lets a 
vehicle stop sending messages for a random period before changing its 
pseudonym. After this period, the vehicle resumes broadcasting beacon messages 
with a new pseudonym. When the silent period is sufficiently long and several 
vehicles were simultaneously silent, an adversary cannot link old and new 
pseudonyms of each vehicle. However, long silent periods negatively affect the 
accuracy of safety applications. On the other hand, Freudiger \etal 
\cite{Freudiger07} proposed placing cryptographic mix-zones (CMIX) in road 
intersections where pseudonyms are forced to be changed in these regions. When 
passing by a mix-zone, vehicles obtain a symmetric key from the RSU and encrypt 
all messages exchanged within this zone. However, the mix-zone suffer from 
several shortcomings due to its fixed placement. Firstly, its effectiveness 
depends on the number of vehicles that enters the zone and changes their 
pseudonyms. Secondly, mix-zones are prone to timing and transition attacks 
where the adversary has knowledge of the probabilities of the exit direction 
and the time spent within the zone for each entering vehicle. Although these 
attacks can reduce the effectiveness of the mix-zone significantly, several 
works considered this problem such as \cite{Palanisamy15}. Thirdly, mix-zones 
are basically depending on RSUs in its operations, although it is not expected 
that RSUs will be widely spread in the initial deployment of VANET. 
%Therefore, it is important to consider the impact of a privacy scheme on 
%safety applications to better understand this trade-off between privacy and 
%safety. 

In this paper, we propose a context-adaptive privacy scheme (CADS) that 
utilizes silent periods to deliver unlinkability among subsequent pseudonyms. 
This scheme is a significant improvement of our recent work, context-aware 
privacy scheme (CAPS) \cite{Emara15b}. The CAPS allows a vehicle to monitor its 
context and choose when to switch to silence and change the pseudonym. 
Additionally, the CADS allows the driver to choose her privacy preference 
whether low, normal or high levels. It optimizes the internal parameters 
dynamically according to the density of the surrounding traffic and the 
driver's privacy preference. It also preserves the vehicle pseudonyms pool for 
a longer time if the pseudonym is already changed with a probable confusion. 
Our contributions in this paper can be summarized as follows:
\begin{itemize}
	\item Propose a context-adaptive and user-centric privacy scheme for VANET 
	(CADS)
	\item Evaluate the gained privacy of CADS against both passive and active 
	adversaries using a well-defined traceability-based metric.
	\item Evaluate the quality of service (QoS) of forward collision warning 
	(FCW) application when applying the CADS. 
\end{itemize}

The rest of the paper is organized as follows. We discuss related work in 
Section \ref{sec:rw}. In Section \ref{sec:method}, we describe the system and 
adversary models, explain how the privacy and the quality of service of a FCW 
application are evaluated, and present the vehicle traces. We explain CAPS 
briefly in Section \ref{sec:caps} while we propose and evaluate CADS in Section 
\ref{sec:cads}. Finally, we show conclusions and future work in Section 
\ref{sec:con}.

\section{Related Work}
\label{sec:rw}
On the one hand, preserving location privacy gained a significant attention in 
the past decade. The silent period scheme was first proposed by Huang \etal 
\cite{Huang05} and applied in VANET by Sampigethaya \etal 
\cite{Sampigethaya05}. Li \etal \cite{Li06} proposed Swing and Swap schemes for 
wireless networks that are based on silent period. In Swing scheme, a node 
changes its identifier and enters silence only when changing its speed and 
direction and there is at least one nearby node. The node broadcasts an update 
message to inform its neighbors which may initiate an update process if their 
privacy is less than the desired. In Swap scheme, nodes can exchange their 
identifiers with probability 0.5 after informing the authentication server. 
Furthermore, Gerlach and Guttler \cite{Gerlach07} propose the concept of 
\textit{mix context} in VANET where a vehicle changes its pseudonym if there 
are N neighbors within a small distance after maintaining the pseudonym for 
stable time. The vehicle assesses the situation after each change to ensure the 
change was successful by measuring if other vehicles changed their pseudonyms 
in the same time step. If this is not the case, the change cycle is repeated. 
Butty\'{a}n \etal \cite{Buttyan09}  propose to stop sending messages when the 
vehicle speed drops to low speeds, for example at intersections. The idea 
behind choosing low speed is that fatal accidents are less likely to occur at 
low speed and places like intersections are natural mix areas where many 
vehicles are located in close proximity. Wei and Chen \cite{Wei10} propose to 
obfuscate position, speed and heading of vehicle within the radius of safe 
distance calculated by safety analysis algorithm. They also propose changing 
the length of the silent period based on the distance from other vehicles. 
Thus, the closer the vehicles are to one each other, the shorter the silent 
period. 

On the other hand, Calandriello \etal \cite{Calandriello07} were one of the 
first to measuring the impact of changing pseudonyms on safety. They evaluated 
the reception timing of the new pseudonym at several distances and relative 
speeds. Lefevre \etal \cite{Lefevre13} analyzed the influence of the duration 
of silent period on the effectiveness of intersection collision avoidance (ICA) 
systems. They proposed an ICA system and evaluated a silent period scheme in 
terms of missed and avoided collisions. They claim that the ICA system can 
function well with silent periods of less than two seconds.

Our proposed scheme is inspired by but more advanced than the previously 
mentioned techniques. Firstly, CADS does not rely on fixed heuristics, such as 
a changing velocity \cite{Buttyan09} or a density threshold \cite{Gerlach07}, 
to identify the adequate mix context to change pseudonyms. The dynamic 
context-based technique of CADS allows short but efficient silence periods so 
that the quality of safety applications is not significantly affected and also 
conserves pseudonyms when vehicle privacy is probably preserved. Secondly, CADS 
considers the driver preferences regarding privacy so that it can maximize the 
privacy level when the driver goes to a sensitive place. 
Thirdly, we employed traceability as a privacy metric rather than the size of 
anonymity set or entropy. The traceability metric expresses on the correctness 
of an adversary to reconstruct vehicle traces from beacons. The 
uncertainty-based metrics, such as entropy, mis-estimate the location privacy 
of users, as shown by Shokri \etal \cite{Shokri11}. Fourthly, we considered the 
trade-off between privacy and safety by measuring the QoS of a FCW application. 
Last but not least, we employed realistic large-scale vehicle traces along with 
a robust vehicle tracker based on multi-target tracking technique in evaluation 
which confirms the scheme practicability, applicability and scalability in 
real-world situations.

\section{Methodology}
\label{sec:method}

\subsection{System Model}
We assume each vehicle is equipped with an on board unit (OBU) which it uses to 
communicate with other vehicles and broadcast \textit{beacon} messages 
periodically (1-10 Hz). The beacon information includes a pseudonym, a 
timestamp and the current vehicle state (i.e., position, speed and heading). 
Vehicles use a state-of-the-art pseudonym issuing process such as 
\cite{Khodaei14} to retrieve a pool of pseudonyms to be used one by one in the 
V2X communication. Pseudonyms have a \textit{minimum pseudonym time} during 
which they must be kept unchanged to ensure stable communication. After that 
time, a vehicle changes the pseudonym according to the adopted privacy scheme. 
The European standard ETSI TS 102 867 recommends changing a pseudonym every 
five minutes \cite{ETSI102867} while the American SAE J2735 standard recommends 
changing it every 120 s or 1 km, whichever comes last \cite{SAEJ2735}. Since 
beacons are essentially used by safety applications, the broadcast information 
has to be as precise as possible. Thus, we assume each vehicle is equipped with 
a GPS receiver and combines the obtained measurements with its internal sensors 
to minimize the position error up to 50 cm. This small error is recommended in 
\cite{Shladover06} and also realized in systems such as \cite{Sengupta06} to be 
able to achieve useful Cooperative Collision Warning applications (CCW). We 
assume that a vehicle maintains the states of its nearby vehicles located 
within its communication range (e.g., 300 m) using a multi-target tracking 
(MTT) algorithm. The utilization of a MTT algorithm for neighbor states 
maintenance is two-fold. First, it allows a vehicle to predict, with the help 
of a Kalman filter, the state of neighbors even if their beacons are delayed or 
missed due to a communication error or a silence period. As a result, the MTT 
algorithm can enhance the effectiveness of safety applications. Second, the MTT 
algorithm supports the vehicle in choosing the appropriate situation to change 
pseudonyms that increases the likelihood of tracker confusion, as will be 
explained in Sections \ref{sec:caps} and \ref{sec:cads}.

\subsection{Adversary Model}
\label{sec:tracker}
We concern protecting vehicles from both (i) a global passive adversary (GPA) 
and (ii) a local active adversary (LAA). The GPA deploys low-cost receivers 
over a large part of the road network and eavesdrops on all exchanged messages. 
Having an external adversary that can cover the whole network may seem 
challenging, but we assume the worst case scenario. Also, this model is 
realizable, for example, by an untrusted service provider through its deployed 
roadside units. The main objective of the adversary is a \textit{tracking 
attack} or reconstructing all vehicle traces from their beacon messages. Thus, 
we assume that the driver's location privacy is determined by the protection 
level against this attack. Although breaching the driver's location privacy 
requires de-anonymization of the reconstructed traces, the de-anonymization 
process is out of the paper scope. However, we assume that the more complete 
and correct the reconstructed traces, the more successful the de-anonymization 
process. 

The adversary achieves its objective by correlating the beacons of a vehicle by 
pseudonym matching. When a vehicle changes its pseudonym, the adversary uses a 
multi-target tracking algorithm, such as NNPDA \cite{Emara13TR}, to correlate 
the messages of the old and new pseudonyms. If the adversary covers only a 
small part of the road network, it can still track vehicles within this limited 
area, but such tracking may not be valuable regarding de-anonymization as it 
does not reflect complete traces. Although powerful adversaries can track 
vehicles using already-deployed cameras spread over the road network, the cost 
and inefficiency of global camera-based attacks will be much higher than those 
for global beacon-based attacks \cite{Freudiger07}.

The LAA can send authenticated messages to the network through a limited amount 
of compromised vehicles driving in the road network. It is assumed that it is 
extremely difficult for an active adversary to be global. The LAA aims mainly 
to deplete the pseudonym pools of the victim vehicles by forcing repeated 
pseudonym changes. If its pool is depleted, the victim will attempt to refill 
its pseudonym pool by initiating a pseudonyms issuing process with a trusted 
service provider, which is not always accessible. This adversary tries to mimic 
conditions that make its surrounding vehicles change their pseudonyms by 
exploiting the procedures of the privacy scheme. Since our proposed scheme 
depends on the vehicle context to change the pseudonym, it is important to 
evaluate active internal adversaries. The encryption-based privacy schemes 
(such as CMIX \cite{Freudiger07}) fails in protecting vehicles from this 
adversary model because the compromised vehicles can obtain symmetric keys from 
RSUs and decrypt all exchanged messages. This gives another advantage for our 
proposed scheme. 

%The problem is that most of safety applications send messages (aka beacons) 
%containing the vehicle position, speed, heading and etc. which can be 
%exploited 
%to know the whereabouts of the driver. 
\subsection{Privacy Evaluation}
\label{sec:privacyeva}
In privacy evaluation, we use the vehicle tracker proposed in \cite{Emara13TR, 
Emara13w} to measure the traceability of vehicles. However, we tuned this 
tracker to accommodate silent periods used by privacy schemes. Originally, the 
tracker holds a vehicle track without update until \textit{time-to-live} time 
steps and deletes it after that time from its tracks list. We included an 
additional parameter of the maximum silence period (\textit{max-silence}) 
allowed by a privacy scheme. The tuned tracker only marks a vehicle track as 
inactive after \textit{time-to-live} time steps and keeps it for additional 
\textit{max-silence} time steps. When the tracker assigns messages of unmatched 
pseudonyms to its tracks list,  only inactive tracks are considered. This 
modification increases the traceability of the tracker since it eliminates 
matching messages of new pseudonyms with unrelated tracks. 

When evaluating a privacy scheme, we alter vehicle traces according to the 
operations of this scheme to generate pseudonymous beacons. To avoid 
synchronization among vehicles, they enter the road network with a pseudonym of 
a random life time ranges from one second up to the minimum pseudonym time. The 
pseudonymous beacons are provided to the tuned tracker to measure the vehicle 
\textit{traceability}. 

\begin{figure}
	\centering
	\includegraphics[width=1\linewidth]{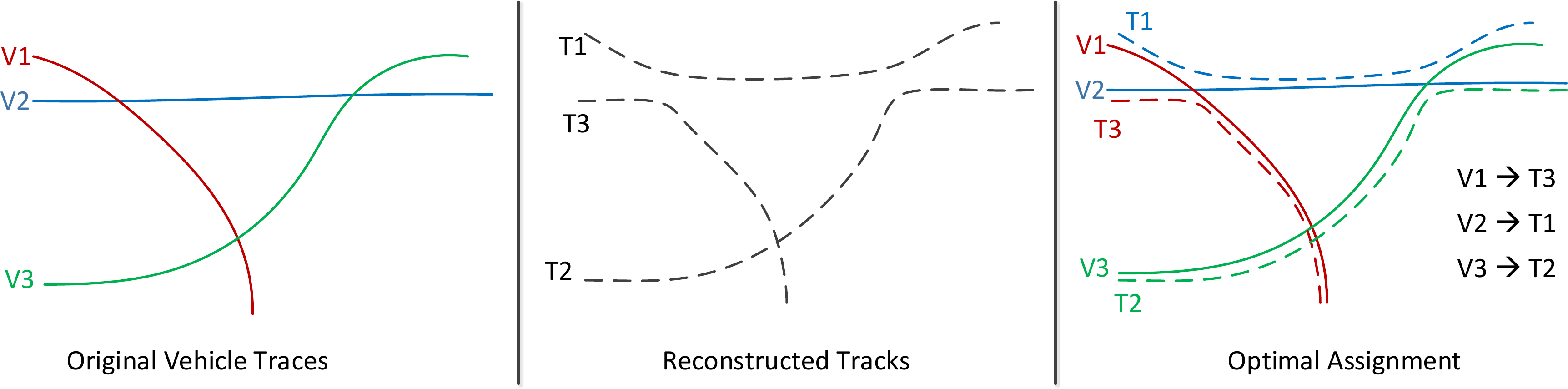}
	\caption{Traceability metric illustration}
	\label{fig:Tracemetric}
\end{figure}

We explain the traceability metric more thoroughly since comparing the 
reconstructed tracks with the original vehicle traces is not trivial, as 
illustrated in Figure \ref{fig:Tracemetric}. In this example, there are three 
traces V1, V2 and V3 (drawn as solid lines on the left) that are reconstructed 
into three tracks T1, T2 and T3 (drawn as dashed lines on the middle). By 
visually comparing both sets, it is clear that each track is reconstructed from 
partial segments of the original traces. For example, T1 is reconstructed from 
segments of V1, V2 and V3. Most of traceability metrics proposed in literature 
\cite{Hoh05, Huang05, Sampigethaya07} may fail to reflect the actual 
traceability level of this adversary. The main issue of their operation is that 
they assign tracks to vehicle traces during the tracking process. In other 
words, they assume the track firstly assigned to a vehicle trace should 
continue with this trace till its end. However, this early assignment 
underestimates the length of the reconstructed tracks. For example, if the 
traceability of V1 is measured by assigning T1 to V1, then this metric shows a 
very short tracking time, although V1 is reasonably reconstructed by T3. 
Therefore, it will be more effective if tracks are assigned to the vehicle 
traces globally after the tracking process is complete.  

The track-to-trace assignment is basically a nonlinear \textit{assignment 
problem} where the total benefit should be maximized. The benefit represents 
the tracking period when a track $t$ assigned to a vehicle trace $v$ 
continuously. Let $l(v, t), \forall v, t \in V, T$ be the maximum continuous 
tracking period when the track $t$ is assigned to the vehicle trace $v$. Note 
that $t$ can be assigned to $v$ for disconnected segments at different times. 
In this case, $l(v, t)$ represents the longest segment. The disconnected 
segments are not summed together because the tracking is discontinued and the 
track may be assigned to another vehicle trace during this discontinuity. The 
adversary cannot reconnect these segments and filter out this wrong assignment 
period because the adversary does not know if he is confused or not. Let 
$\tau_{v}$ be the maximal tracking period of $v$; and it can be obtained by 
solving the following assignment problem: 
\begin{align}
	\label{eq:assignprogram}
	\text{maximize } & \sum_{v \in V} \tau_{v} \nonumber \\
	\text{subject to } & \tau_{v} = \sum_{t \in T} l(v, t) \cdot a_{v, t} , 
	\quad a_{v, t} \in \{0, 1\}, \\
	&\sum_{v \in V} a_{v, t} \leq 1 \quad \forall t \in T \quad and \nonumber  
	\\
	&\sum_{t \in T} a_{v, t} \leq 1 \quad \forall v \in V.\nonumber 
\end{align}
Here, $a_{v, t}$ is the assignment function which equals one if the track $t$ 
should be assigned to the vehicle trace $v$ and equals zero otherwise. Note 
that not all tracks must be assigned to a vehicle trace because the number of 
tracks can be greater than the number of vehicle traces as some tracks are 
reconstructed from partial vehicle traces. Also, not all vehicle traces must be 
assigned to a track because its $l(v, t)$ may not contribute to the maximal 
$\sum_{v \in V} \tau_{v}$. In this case, $\tau_{v}$ equals zero. This 
assignment problem is solved using an auction algorithm considering tracks as 
the bidders, vehicle traces as the items and $l(v, t)$ as the bidding price. 
After the optimal assignment is obtained, the traceability of the whole 
scenario is calculated by counting the percentage of significantly tracked 
vehicles. Thus, the traceability metric $\Pi$ is defined as:
\begin{align}
	\label{eq:traceabilitymetric}
	\Pi = \frac{1}{N} \sum_{v \in V} \lambda_{v} \times 100, \quad
	\lambda_{v} =\begin{cases}1 & \frac{\tau_{v}}{L(v)} \geq 0.90 \\0 & 
	otherwise\end{cases} 
\end{align}
where $L(v)$ is the lifetime of $v$ and $N$ is the total number of traces 
included in the dataset. This metric allows few confusions around the endpoints 
of a vehicle trace (10\% of the trace lifetime) since inaccuracies in endpoints 
can be smoothed by a clustering technique in a re-identification process, as 
shown in \cite{Hoh06}. According to this definition, the privacy of the driver 
is considered breached if the adversary is able to continuously track 90\% of 
her trace. Also, this metric reflects the probability of being tracked by 
calculating the ratio of tracked vehicles rather than how long a tracker can 
estimate from the actual trace as in \cite{Emara13w, Wiedersheim10}.

In some cases, some vehicles never change their pseudonyms during their 
lifetime. Thus, we additionally calculate the \textit{normalized traceability} 
$\Pi_{n}$ by excluding these vehicles since they are easily tracked by the 
adversary and reflected in the original traceability metric. This normalized 
metric considers the effectiveness of the privacy scheme when a vehicle changes 
its pseudonym at least once and can be defined as:
\begin{align} 
	\Pi_{n} &= \frac{1}{N} \sum_{v \in V} \lambda_{v}^{norm} \times 100,\\
	\lambda_{v}^{norm} &=\begin{cases}1 & \frac{\tau_{v}}{L(v)} \geq 0.90 
	\wedge psd(v, k_0) \neq psd(v, k_0+L(v))\\0 & otherwise\end{cases} 
\end{align}
where $psd(v, k_0)$ and $psd(v, k_0+L(v))$ are the pseudonyms of the vehicle 
$v$ at the first and last time steps in its lifetime, respectively.
\subsection{Quality of Safety Application}
It is important to evaluate the impact of a privacy scheme on safety 
applications because these applications require accurate and continuous 
information about nearby vehicles but, in the same time, privacy schemes 
usually perturb such information. We use our methodology proposed in 
\cite{Emara15} to evaluate the impact of a privacy scheme on a forward 
collision warning (FCW) application. In this method, vehicles estimate the 
states of the nearby vehicles when applying the evaluated privacy scheme and 
calculate the error $\delta$ between their estimation and the ground truth. 
Then, the probability of correctly calculating the main application factors is 
estimated using these error samples in Monte Carlo analysis. The main factors 
of the FCW application are (1) identifying the correct lane of the other 
vehicle and (2) calculating the time to collision (TTC) accurately within small 
tolerance (e.g., within 500 ms). 
\begin{figure}
	\centering
	\includegraphics[scale=.6]{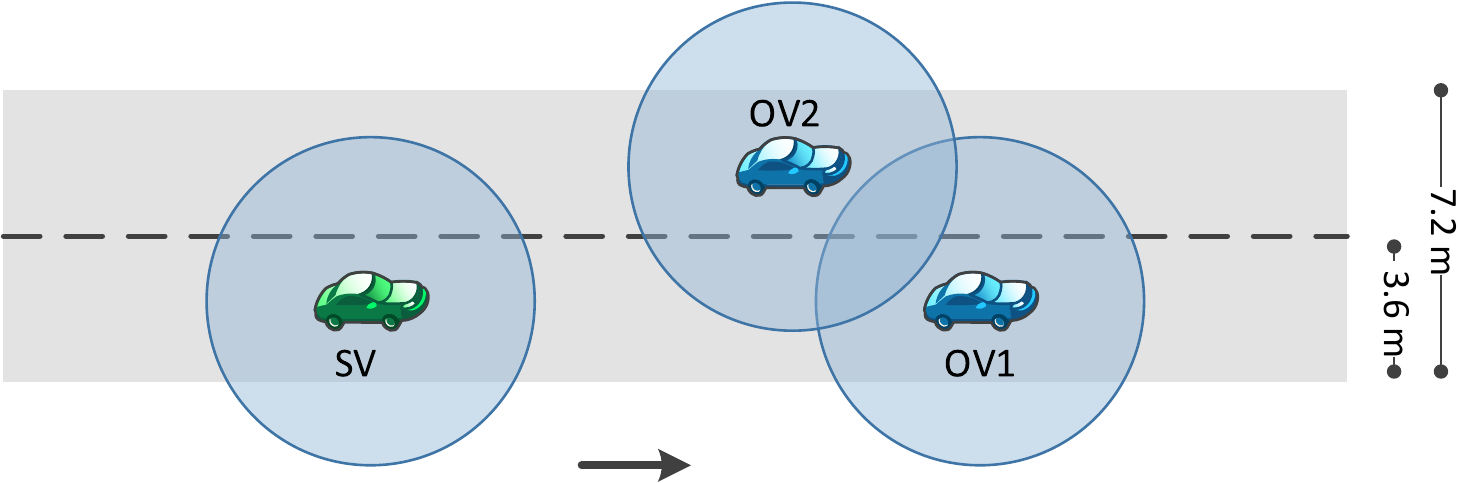}  	
	\caption{Forward collision warning scenario.}
	\label{fig:fcw}
\end{figure}

For the first application requirement, the SV must correctly identify that OV1 
is in its own path (i.e., high sensitivity) while OV2 is not (i.e., high 
specificity), as shown in Figure \ref{fig:fcw}. The criteria for identifying an 
OV as in path is that its lateral position is within $\pm$ 1.8 m of the lateral 
position of the SV, assuming a 3.6 m lane width. Otherwise, it should be 
identified as not in path. In our analysis, we set the true lateral position of 
the SV as same as the lateral position of OV1, while the true position of the 
OV2 is located in the center of the next lane. Thus, the measured lateral 
positions $y$ of SV, OV1 and OV2 are obtained by adding the errors to their 
true positions as follows: 
\begin{equation}
\begin{array}{lcl} 
y_{SV} &=& 1.8 + \mathcal{N}(0, 0.5) \\
y_{OV1} &=& 1.8 + \delta y \\
y_{OV2} &=& 5.4 + \delta y 
\end{array}
\end{equation}

where $\delta y$ is an error sample in the lateral position. Therefore, the 
true and false positive probabilities for correctly identifying lanes of the 
OVs can be calculated by:
\begin{align} 
	P_{true+} &= P(|y_{OV1} - y_{SV}| \leq 1.8) \\
	P_{false+} &= P(|y_{OV2} - y_{SV}| \leq 1.8) 
\end{align}

For the second requirement, we assume that the SV is approaching the OV1 at 
speed differences $\Delta s$ of 5 m/s and 15 m/s. The assumed true TTC is set 
to three seconds as an example; thus, the true position of OV1 is generated to 
be three seconds distance from the true position of SV and is calculated based 
on the evaluated speed difference as follows: 
\begin{equation}
\begin{array}{lcl} 
x_{SV} &=& \mathcal{N}(0, 0.5) \\
x_{OV1} &=& 3 \cdot \Delta s + \delta x \\
\dot{x}_{SV} &=&  \hat{x}_{OV1} + \Delta s + \mathcal{N}(0, 0.02 \cdot 
(\hat{x}_{OV1} + \Delta s))\\
\dot{x}_{OV1} &=& \hat{x}_{OV1} + \delta \dot{x}
\end{array}
\end{equation}
where $\hat{x}_{OV1}$ is the true longitudinal speed of the OV1 and $\delta 
\dot{x}$ is an error sample in the longitudinal speed. Here, there is no binary 
classification to calculate false positives; instead, we calculate the 
probability of calculating TTC within a small tolerance of 500 ms. This 500 ms 
tolerance is chosen by Shladover and Tan \cite{Shladover06} as the maximum 
tolerance for issuing a useful warning. They also analyzed the implication of a 
desirable tolerance of 200 ms but they found that it requires a positioning 
accuracy of 20 cm to attain this restrict tolerance, wherefore we considered 
only the maximum tolerance of 500 ms. Therefore, the TTC and the probability of 
correctly estimating it within 500 ms can be calculated by:
\begin{align}
	TTC &= \frac{x_{OV1} - x_{SV}}{\dot{x}_{SV} - \dot{x}_{OV1}} \\
	P_{TTC} &= P(|TTC - 3| \leq 0.5) 
\end{align}
In this equation, we determine how frequently the difference between the 
calculated TTC and the true TTC (3 s) is less than the tolerance threshold of 
0.5 s. Finally, the probability of the FCW application ($P_{FCW}$) can be 
obtained by multiplying all three probabilities together, assuming they are 
independent, as follows:  
\begin{align}
	P_{FCW \Delta s} &= P_{true+} \times (1 - P_{false+}) \times P_{TTC \Delta 
	s} 
\end{align}

Experiments show that estimating TTC in high speed differences is much more 
accurate than low speed differences with the same position noise. Therefore, 
the QoS of the FCW application ($QoS_{FCW}$) is defined as $P_{FCW \Delta s = 
5}$ multiplied by 100 to obtain a percentage, as follows:
\begin{align}
	QoS_{FCW} &= P_{FCW \Delta s = 5} \times 100
\end{align}

\subsection{Vehicle Traces}
\label{sec:vtrace}
We employ realistic vehicle traces in evaluation which are obtained from 
\cite{VtraceWeb11}. This dataset is mainly based on the data made available by 
the TAPASCologne project \cite{TAPASCologne10} which is an initiative by the 
Institute of Transportation Systems at the German Aerospace Center (ITS-DLR). 
This dataset reproduces vehicle traffic in the greater urban area of the city 
of Cologne, Germany with the highest level of realism possible. The street 
layout of the Cologne urban area is obtained from the OpenStreetMap (OSM) 
database. The microscopic mobility of vehicles is simulated using the 
Simulation of Urban Mobility (SUMO). The source and destination of vehicle 
traces are derived through the Travel and Activity PAtterns Simulation (TAPAS) 
methodology. Uppoor \etal \cite{Uppoor2014} pointed out several problems when 
combining these data sources to produce traffic data. Among these problems, 
vehicles are moving rapidly to large traffic jams, travel times are unrealistic 
and vehicle speeds turn to very low values. Uppoor \etal resolved these 
problems so that the synthetic traffic match that observed in the real world, 
through real-time traffic information services. This is why we name this 
dataset as realistic traces.

\begin{figure}
	\centering
	\includegraphics[scale=.15]{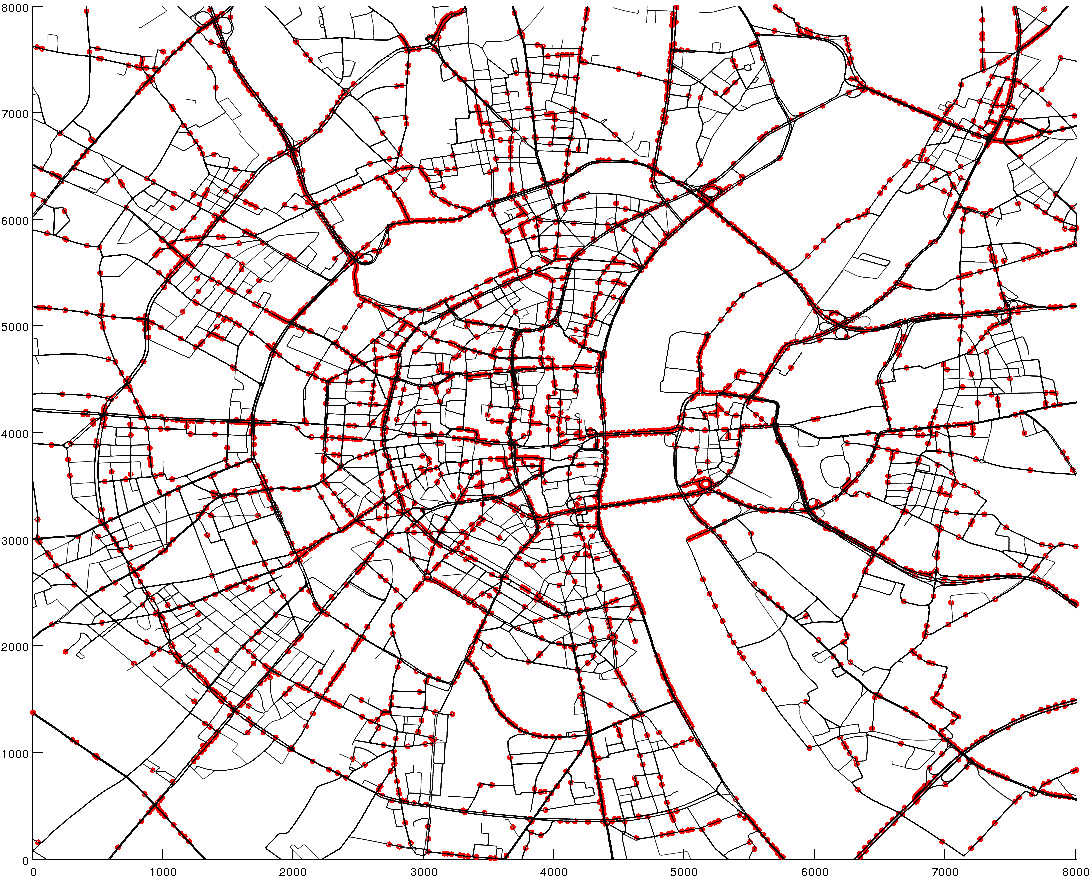} 
	\caption{The road map of the realistic traces. }	
	\label{fig:mapskoln}
\end{figure}	 

We obtained the two-hour sample published online \cite{VtraceWeb11} and 
selected 30 min (from 6:15 AM till 6:45 AM) for the middle 64 $km^2$ region, as 
shown in Figure \ref{fig:mapskoln}. We selected this time period because the 
vehicle density increases dramatically, which provides a challenging evaluation 
for the operation of privacy scheme in different densities. As we cropped the 
vehicle traces in both space and time, we excluded very short traces that move 
within 100 $m^2$ or start and end in less than 15 s. There are 19,704 remaining 
traces with increasing density, ranging from 1,929 to 4,572 simultaneous 
vehicles in the first and last time steps, respectively. The vehicle positions 
in the last time step are drawn as red spots in Figure \ref{fig:mapskoln}. 
Moreover, we processed the dataset to calculate the heading and velocity in the 
$xy$-coordinates using every two consecutive time steps for each vehicle. The 
last heading value is preserved when the vehicle stops and is changed when it 
starts to move.

\section{Context-aware Privacy Scheme (CAPS)}
\label{sec:caps}
The basic concept of our Context-aware Privacy Scheme (CAPS) is to determine 
the appropriate context in which a vehicle should change its pseudonym. This 
approach aims to increase the effectiveness of such changes against tracking 
and avoid wasting pseudonyms in easily traceable situations. More specifically, 
a vehicle continuously monitors other vehicles located within its communication 
range and tracks their beacons using an NNPDA tracker. As explained in our work 
\cite{Emara13TR}, the NNPDA is an efficient multi-target tracking algorithm 
that has exhibited a high tracking accuracy for anonymous beacons with 
different amounts of noise and beaconing rates. 

\begin{figure}[h]
	\centering
	\includegraphics[width=.6\linewidth]{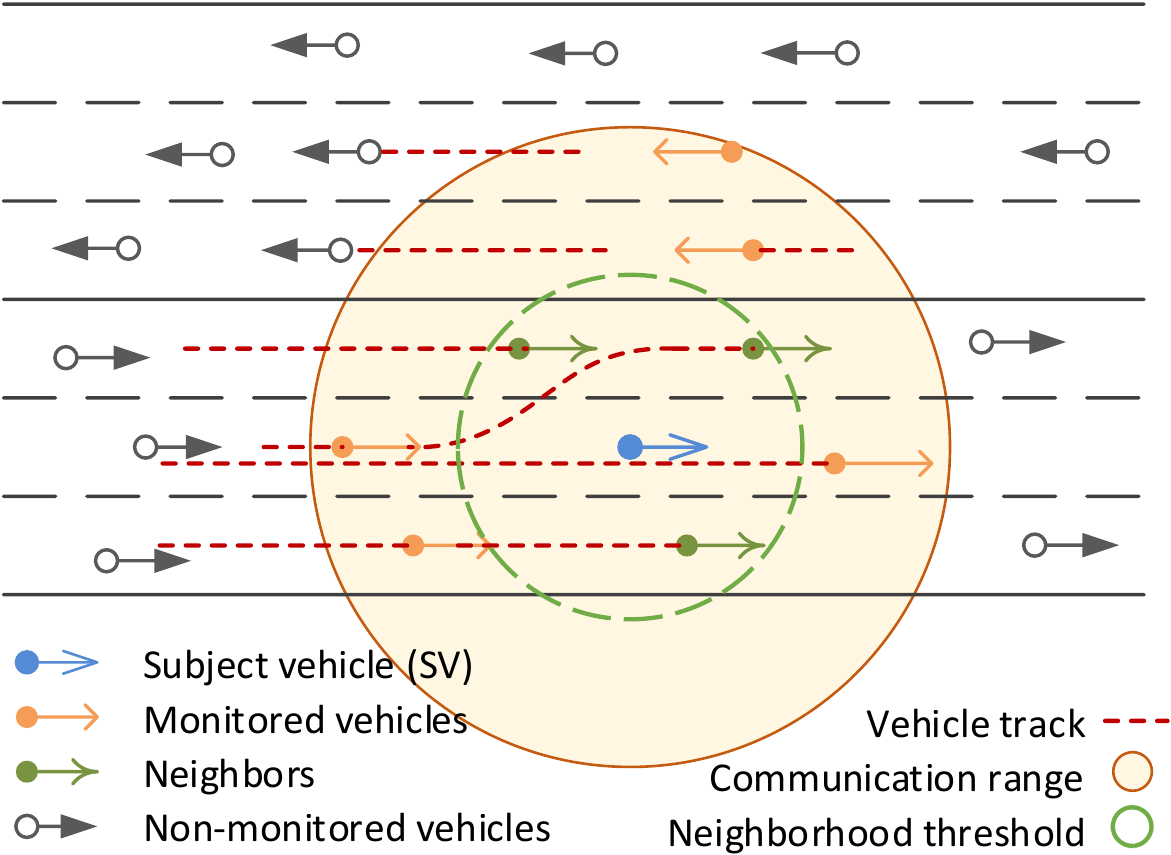}
	\caption{Illustration for the CAPS operations}
	\label{fig:caps}
\end{figure}

As illustrated in Figure \ref{fig:caps}, the CAPS works as follows. During its 
active status, the subject vehicle (SV) uses its current pseudonym in beacons 
until the pseudonym lifetime reaches a minimum time. Once it exceeds this time, 
the vehicle checks whether any of monitored \textit{neighbors} missed its 
beacons for several time steps. Here, neighbors refer to vehicles located 
within a predefined radius from the subject vehicle (e.g., 50 or 100 m). If the 
SV finds a silent neighbor, it turns to silence as well. Otherwise, it 
continues using its current pseudonym until its lifetime reaches a maximum 
pseudonym time and then the vehicle turns to silence. 

\begin{figure}[h]
	\centering
	\includegraphics[width=.7\linewidth]{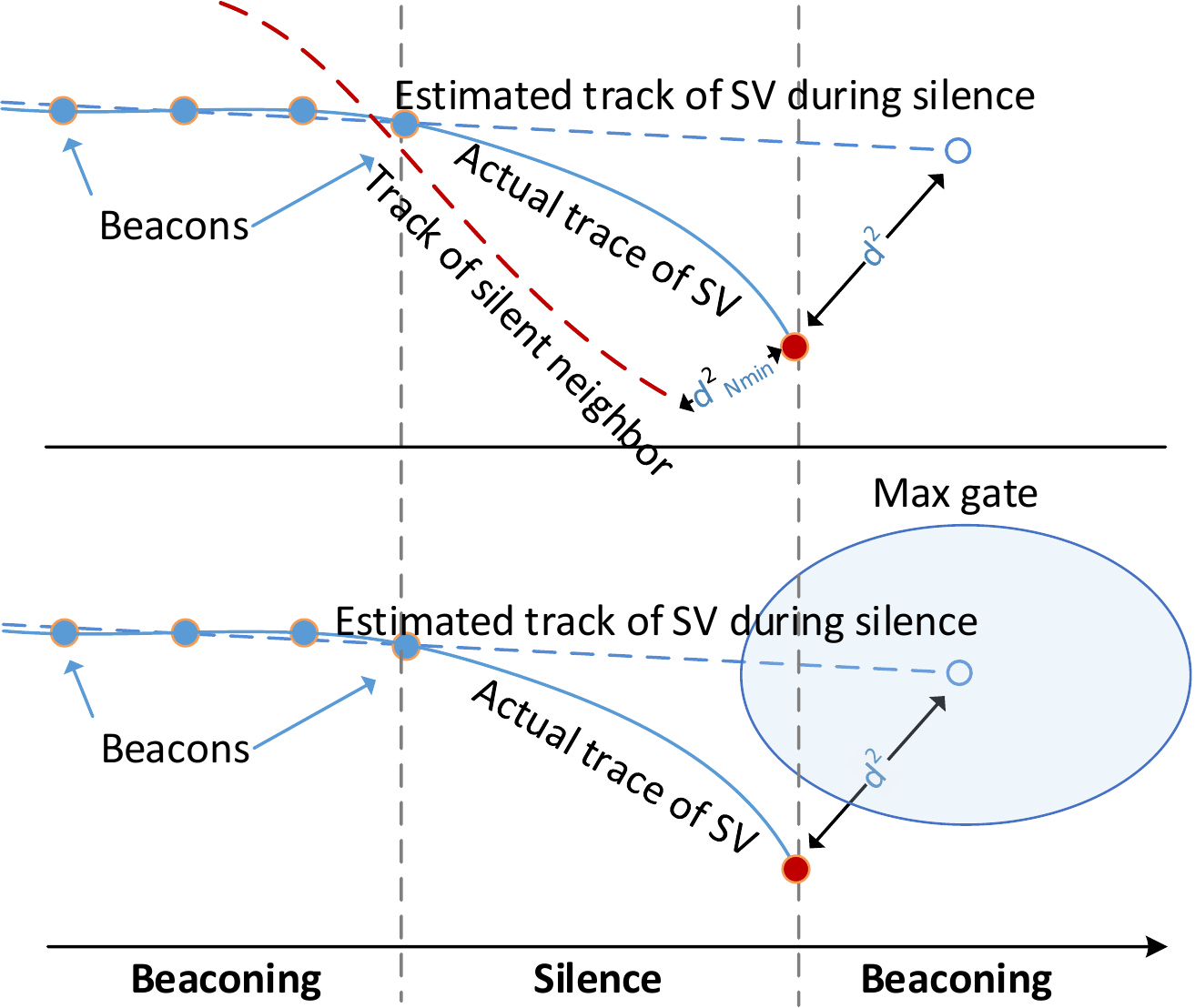}
	\caption{Illustration for the two conditions to exit silence.}
	\label{fig:capscond}
\end{figure}

When a vehicle is silent, it returns to beaconing under more complex conditions 
based on the \textit{gating} phase of vehicle tracking. It was explained in 
\cite{Emara13TR} that a gating process is required in target tracking to 
eliminate unlikely measurement-to-track associations from being tested. It 
requires any new measurement to be located within the track \textit{gate} to be 
a valid candidate for association with this track. The most common gating 
technique is ellipsoidal which defines the norm of the residual vector 
($d^{2}$):
\begin{equation}
d^{2} = \tilde{z}^{T} S^{-1} \tilde{z}
\end{equation}
where $\tilde{z}$ and $S$ are the residual vector and its covariance matrix 
obtained from the Kalman filter, respectively. We exploit this fact and require 
the beacon after silence to achieve one of the following two conditions to 
guarantee no correlation with previous beacons. As illustrated in Figure 
\ref{fig:capscond}, the SV state should be nearer to the track of a silent 
neighbor than its original track or completely outside the gate of its original 
track. When these conditions hold, the adversary will most probably become 
confused when attempting to correlate this new beacon because it will not be 
assigned to its original track.

Formally, when the SV is silent, it continues monitoring surrounding vehicles 
and waits for the minimum silence time. Once exceeded, it checks if one of the 
following conditions holds regarding the norm of the residual vector ($d^{2}$) 
between its actual and estimated states: 

\begin{enumerate}
	\item $d^{2} > d^{2}_{Nmin}$, where $d^{2}_{Nmin}$ is the minimum norm of 
	the residual vector between the SV actual state and the estimated states of 
	its silent neighbors, as shown in the upper part of Figure 
	\ref{fig:capscond}. 
	\item $d^{2} > max\_gate$, where $max\_gate$ is the maximum gate that the 
	adversary may use, as shown in the lower part of Figure \ref{fig:capscond}. 
\end{enumerate}

If one of these conditions holds, this new beacon is likely to be mixed with 
one of its silent neighbors or recognized as a new vehicle. Therefore, it is a 
suitable time to exit silence with a new pseudonym. If these conditions never 
occur, the SV remains silent until a maximum silence time is reached. 

\subsection{CAPS Evaluation}
\begin{figure*}
	\centering
	\subfigure[Traceability]
	{\includegraphics[scale=.45]{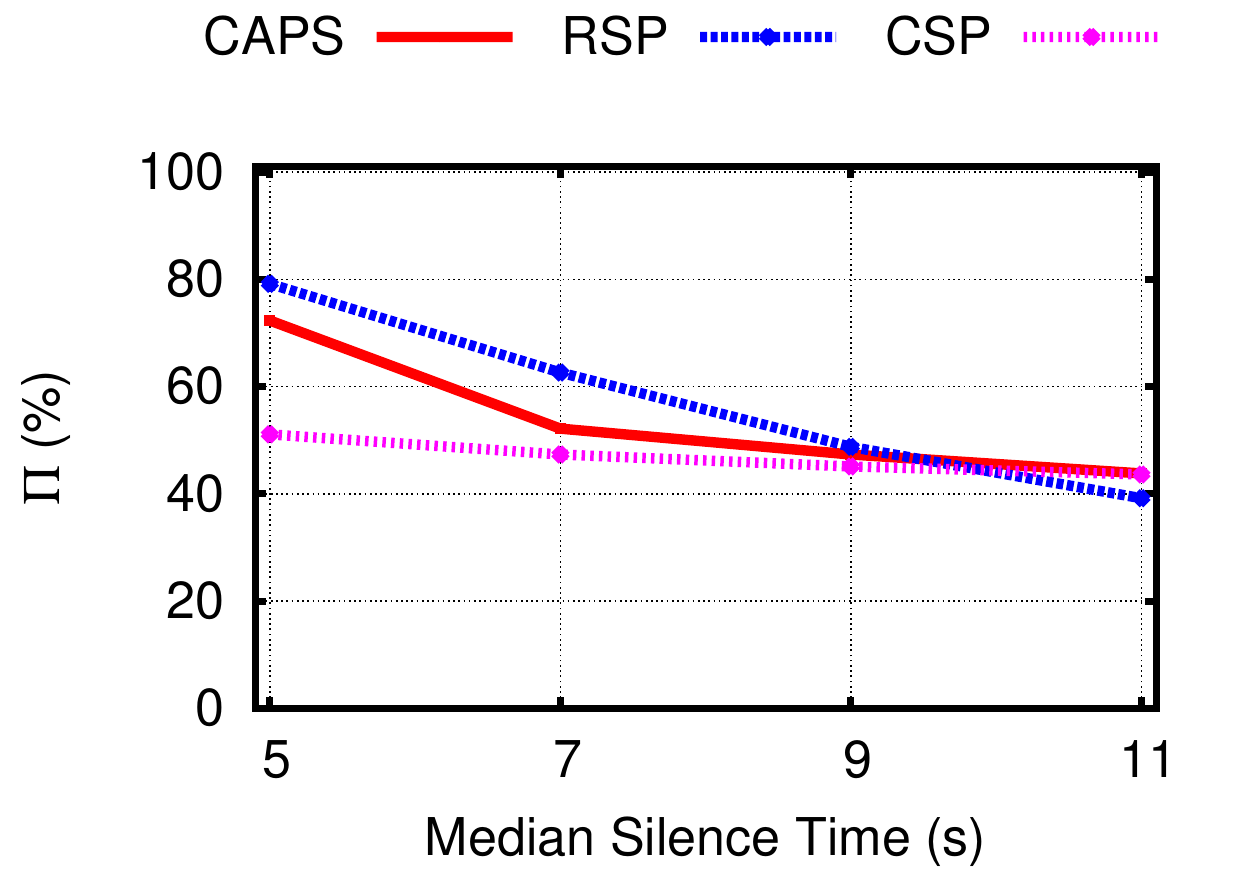} \label{fig:cmpTrace}}
	\subfigure[Normalized traceability]
	{\includegraphics[scale=.45]{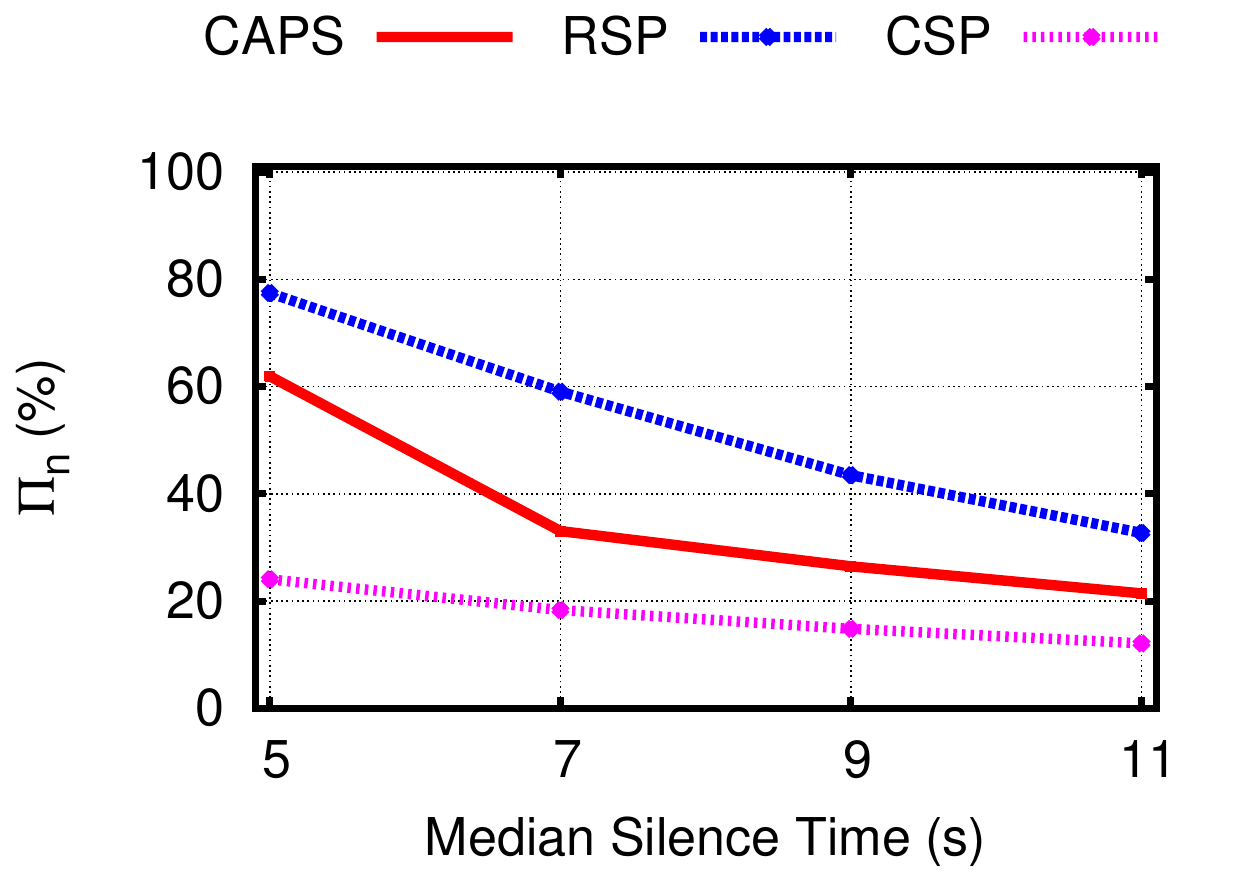} \label{fig:cmpNTrace}}
	\subfigure[QoS ]
	{\includegraphics[scale=.45]{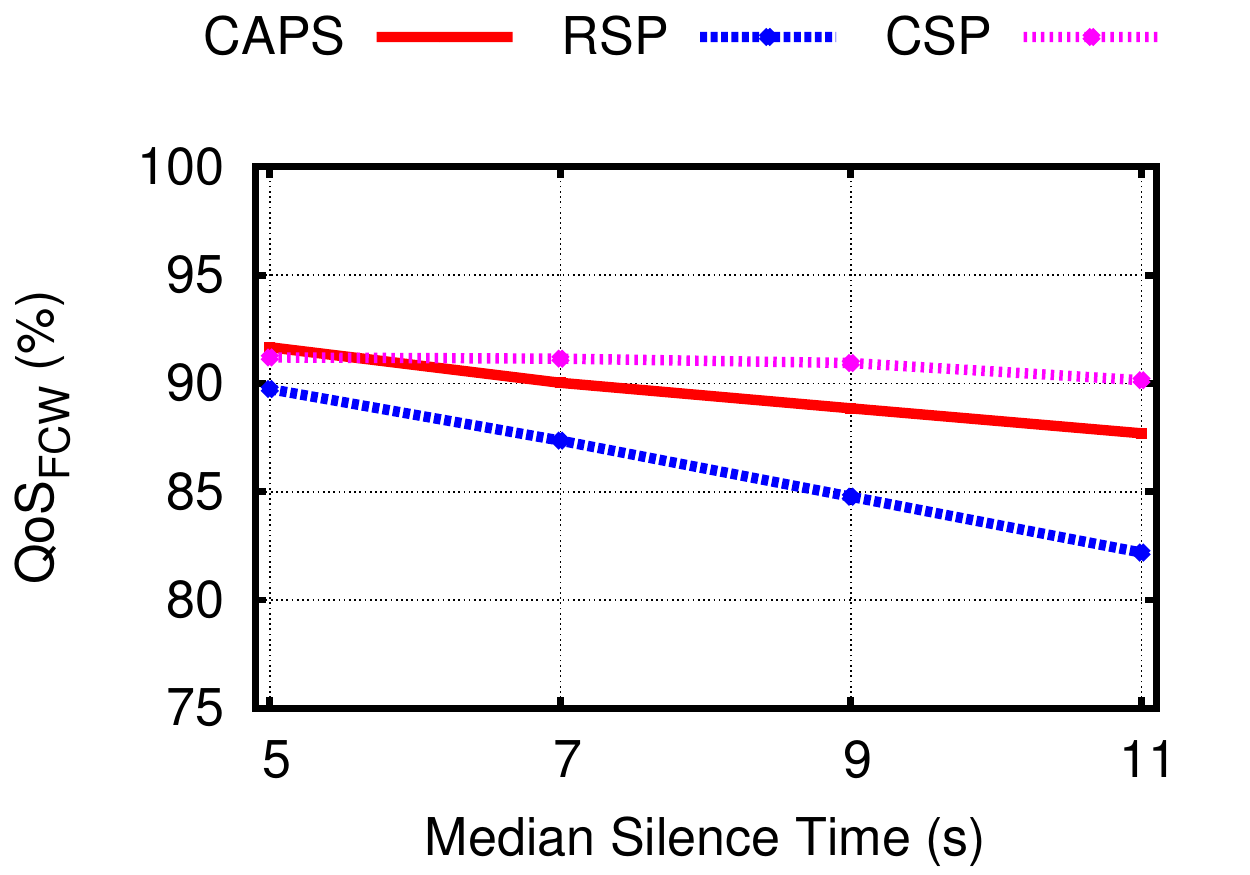} \label{fig:cmpQos}}
	\caption{Evaluation comparison among CAPS, RSP and CSP.}
	\label{fig:capsres}
\end{figure*}

We evaluate the CAPS in comparison with two silent-based schemes: the random 
and coordinated silent period schemes. The random silent period (RSP) allows a 
vehicle to change its pseudonym after a fixed pseudonym time and keep silent 
for a uniformly random period within a preset range (e.g., from 3 to 13 s). As 
the two schemes have different assumptions and parameters, they are aligned 
based on the median silent and pseudonym times for all vehicles, actually 
performed in the simulation. The maximum pseudonym time of CAPS is 300 s while 
the fixed pseudonym time of RSP is 120 s. In CAPS, we assigned 50 m to the 
neighborhood threshold and 60 s to the minimum pseudonym time. The coordinated 
silent period (CSP) is proposed by Tomandl \etal \cite{Tomandl12} in their 
comparison of silent period and mix zone schemes. CSP coordinates all vehicles 
in the network to remain silent and change pseudonyms synchronously. CSP seems 
to be theoretical since the coordination overhead in real world situations 
increases dramatically \cite{Tomandl12}. However, CSP increases the privacy 
significantly because it maximizes the size of the anonymity set at every 
pseudonym change. The pseudonym lifetime of CSP is 300 s and all vehicles are 
synchronized so that they turn to silence and change pseudonyms simultaneously.

In Figure \ref{fig:capsres}, we show the traceability $\Pi$, normalized 
traceability $\Pi_n$ and QoS of a FCW application for all schemes. The CAPS 
slightly reduces the traceability $\Pi$ especially in short median silence 
times (up to 10\% lower than RSP) as shown in Figure \ref{fig:cmpTrace}. In 
longer silent times, all schemes result in similar traceability. Since many 
vehicles did not change their pseudonyms in simulation, CAPS and CSP reduced 
the normalized traceability by up to 20\% and 30\% from the traceability 
metric, respectively. This happens because the CAPS chooses the right context 
to change the pseudonym which increases the probability of tracker confusion. 
Hence, the pseudonym change made by the CAPS is much more effective than that 
made randomly by the RSP. For CSP, vehicles are synchronized which maximizes 
the size of the anonymity set and increases the tracker confusion 
significantly.  

For the QoS, the CAPS achieves a higher QoS than the RSP (up to 6\% higher) and 
slightly lower than CSP (up to 2\% lower). This QoS decrease in the RSP occurs 
due to the unnecessary and ineffective pseudonym changes. These pseudonym 
changes are followed by silence periods which affects identifying the 
application requirements correctly especially with relatively long silence 
periods. 

\subsection{CAPS Shortcomings}
We note three shortcomings of the CAPS. First, we observe that some vehicles 
change pseudonyms unnecessarily several times with no significant advantage in 
decreasing the traceability. Having a few confusions per trace is sufficient to 
avoid continuous vehicle tracking. However, frequent pseudonym changes and 
confusions may negatively affect the QoS of a safety application, as neighbors 
cannot estimate the vehicle state correctly. Therefore, we propose increasing 
the minimum pseudonym time each time a vehicle changes its pseudonym with a 
probable confusion. Second, the CAPS takes several parameters that may not be 
optimized for different traffic densities and situations. For example, a wide 
neighborhood threshold may be more suitable for sparse traffic than dense 
traffic. Third, the CAPS does not consider the driver's preference regarding 
privacy. In fact, privacy depends on the preferences of the user and the 
technical solutions should be adaptable to empower users to determine what is 
allowed with their personal information \cite{PRECIOSA09}. For example, it may 
be desirable to maximize the privacy level when the driver goes to a sensitive 
place. For these reasons, we propose a more advanced scheme that considers 
these shortcomings, which we call the context-adaptive privacy scheme (CADS) as 
explained next.

\section{Context-Adaptive Privacy Scheme (CADS)}
\label{sec:cads}
The CADS allows a driver to choose among privacy preferences, whether low, 
normal or high. It optimizes the internal parameters dynamically according to 
the density of the surrounding traffic and the driver's privacy preference. In 
addition, it preserves the vehicle pseudonyms pool for a longer time if the 
pseudonym is already changed with a probable confusion. 

\begin{table*} 
	\centering
	\caption{Optimized CADS parameters and their results}	
	\label{tab:cadsparam}
	\begin{tabular}{lccc|ccc}
		\toprule
		&  \multicolumn{6}{c}{Privacy Preference} \\
		&  \multicolumn{3}{c}{Sparse sub-dataset} &  \multicolumn{3}{c}{Dense 
		sub-dataset}\\
		Parameter/Result & Low & Normal & High & Low & Normal & High \\ \hline	
		Max pseudonym time (s)& 240 &  300 &  180  & 240 & 180 & 180 \\  
		Max silence time (s) & 11 &  11 &  11 & 11 &  13 &  11 \\  
		Pseudonym time increment (s) & 60 &  60 &  0 & 60 &  60 &  0 \\  
		Neighborhood threshold (m) &  50 & 100 &  100 & 50 &  50 &  100\\ 
		\hline	
		Traceability (\%)  & 75 &  59 & 49 & 68 &  49 &  38\\ 
		Normalized Traceability (\%) & 52 &  33 & 31 & 43 &  26 & 21\\  
		QoS (\%) 				& 90 &  87 &  85 & 91 &  88 &  85\\ \bottomrule
	\end{tabular}	
\end{table*}  

To optimize the scheme parameters with respect to the surrounding traffic, we 
investigate the performance of the CAPS in two different densities; sparse and 
dense traffic. First, we select two relatively short sub-datasets from the 
realistic vehicle traces with low and high traffic densities. We then test the 
CAPS on each sub-dataset with many parameter combinations and obtain the 
resulting traceability and QoS metrics. Second, to incorporate the privacy 
preference in CADS, we divide the results of the sub-dataset experiments into 
three categories according to the achievable traceability. Next, we identify 
the parameters that result in the best compromise between traceability and QoS 
in each category. Third, we insert these categorized parameters of each density 
into CADS and bind them according to the real-time vehicle density and the 
input privacy preference. 

\begin{figure*}
	\centering
	\subfigure[]
	{\includegraphics[scale=.25]{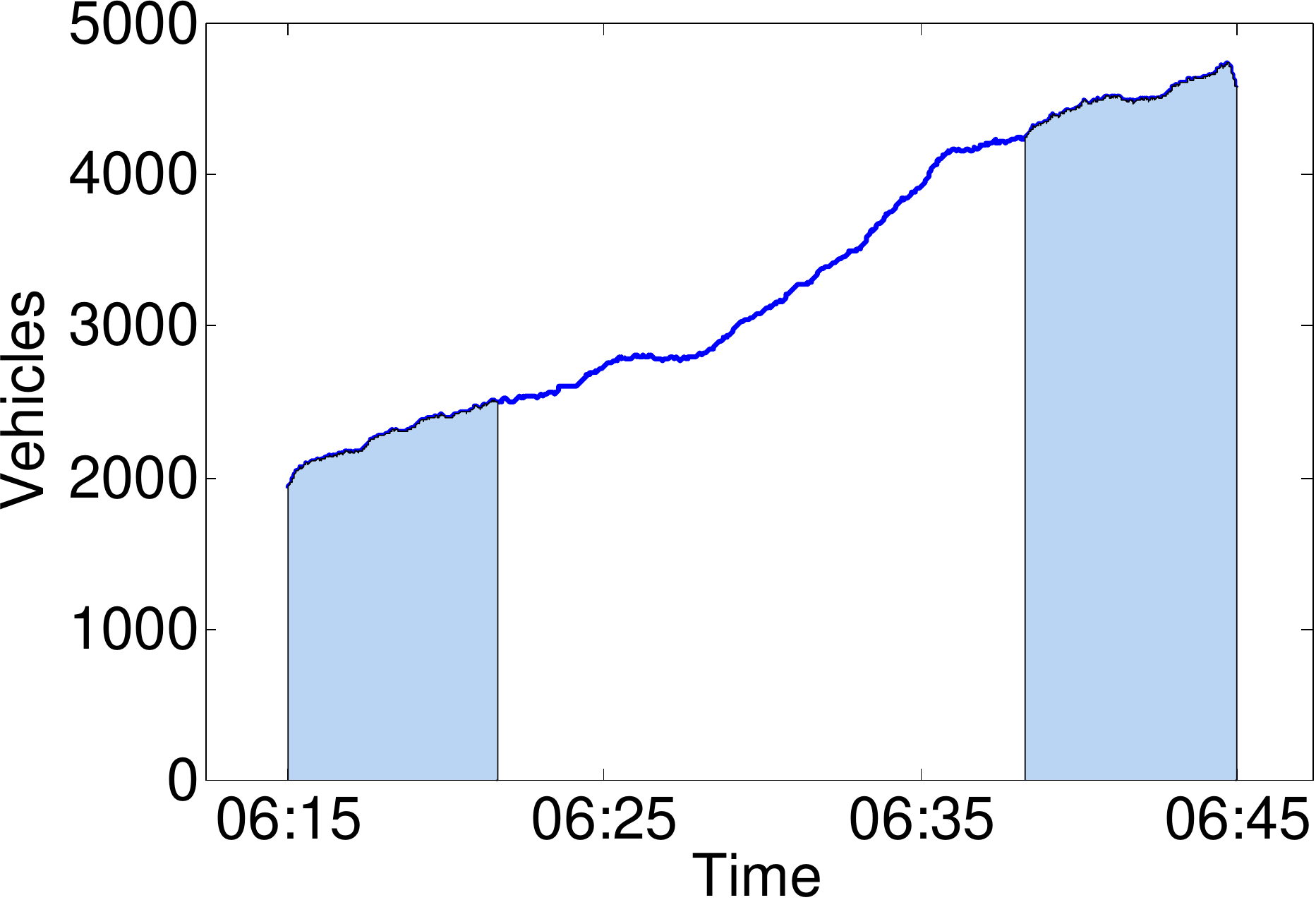} 
	\label{fig:vdensity}}
	\subfigure[Sparse sub-dataset]
	{\includegraphics[scale=.2]{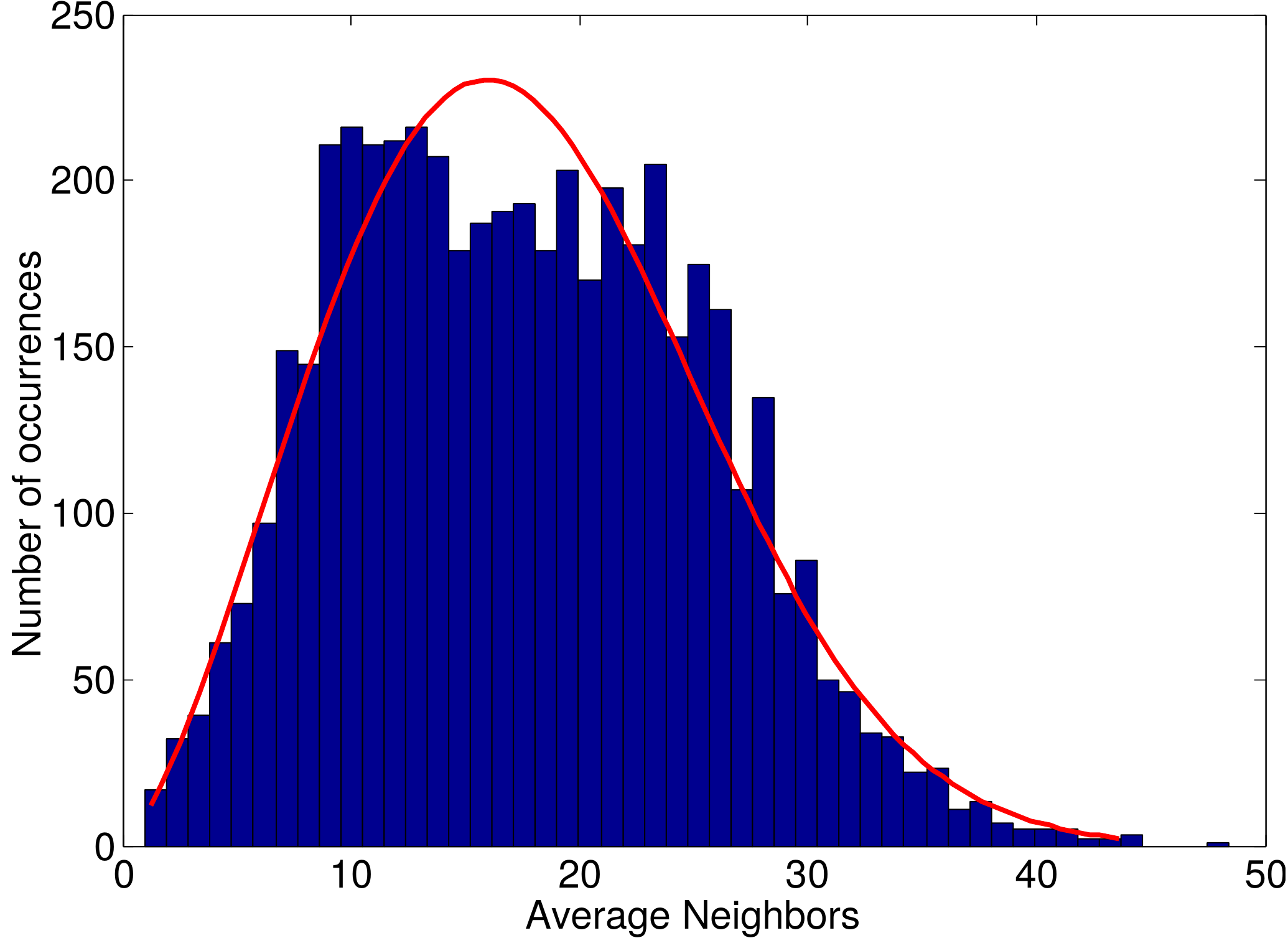} 
	\label{fig:neighborl}}
	\subfigure[Dense sub-dataset]
	{\includegraphics[scale=.2]{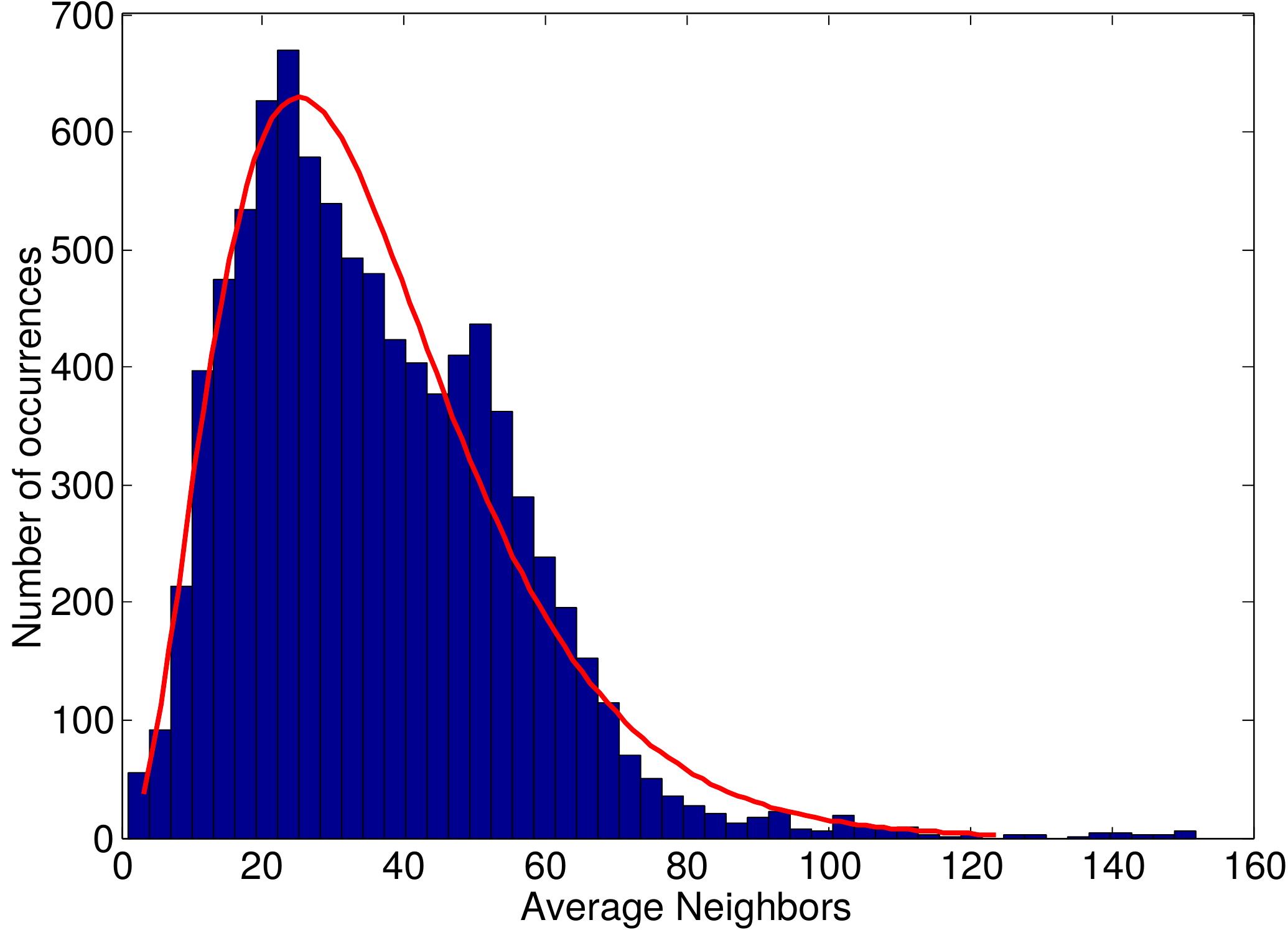} 
	\label{fig:neighborh}}
	\caption{(a) Vehicle density versus time with sub-datasets highlighted. (b) 
	and (c) Average number of neighbors encountered by a vehicle in each 
	sub-dataset.}
	\label{fig:subdataset}
\end{figure*}
\subsection{Sub-datasets Evaluation}
As explained in Section \ref{sec:vtrace}, the vehicle traces have an increasing 
density ranging from 1,929 to 4,572 vehicles. We selected two sub-datasets, 6 
min long each from the beginning and end of vehicle traces, as shaded in Figure 
\ref{fig:vdensity}. We excluded traces that last less than one minute from 
these sub-datasets. The CAPS is then evaluated using each sub-dataset and the 
following parameter combinations: maximum pseudonym times of 180, 240 and 300 
s, maximum silence times of 7, 9, 11 and 13 s, neighborhood thresholds of 50 
and 100 m and increments of the minimum pseudonym time after a probable 
confusion of 0 or 60 s. We run the CAPS using these parameter combinations on 
both sub-datasets and obtain the achieved privacy and QoS metrics. 

\subsection{Parameters Selection}
From all experiments tested in the previous step, we exclude those results with 
a QoS less than 85\% as we assume that the safety application will not operate 
with an acceptable accuracy in such cases. Although the traceability and the 
QoS are proportional, we notice that the QoS varies much less than the 
traceability. Therefore, the results are categorized based on the QoS instead, 
to facilitate categorization. The results are divided into low, normal and high 
privacy levels when they achieve the maximum, average and minimum QoS, 
respectively in each sub-dataset. Thus, the parameters for a high privacy 
preference are selected when a QoS of 85\% is attained. The parameters for a 
low privacy preference are selected when the highest QoS is obtained but with a 
traceability of at most 75\%. This traceability constraint is added to ensure 
some privacy even when low privacy preference is selected. The parameters for 
normal privacy preference are selected when the average QoS is attained with 
the lowest traceability.

In Table \ref{tab:cadsparam}, we show the selected parameter set for each 
privacy preference and vehicle density. In the last three rows, we include the 
resulting traceability and QoS of each parameter set when applied to the 
sub-datasets. We notice that the achievable traceability in the sparse 
sub-dataset is higher than that achievable in the dense sub-dataset. The 
traceability can be decreased using more restrict parameters but only at the 
cost of the QoS. 
\subsection{CADS Algorithm}
The parameter table \ref{tab:cadsparam} is integrated into the CADS to let a 
vehicle choose the adequate parameter set based on the driver's privacy 
preference and the real-time density of the surrounding traffic. A vehicle can 
estimate the traffic density by evaluating the average number of neighbors 
encountered over time. For this purpose, we analyzed the distribution of 
neighbors in both sub-datasets, as shown in Figure \ref{fig:subdataset}. We 
notice that the average number of neighbors that a vehicle encounters is 30 and 
68 with 95\% confidence in the sparse and dense sub-datasets, respectively. 
Therefore, a neighbors threshold of 30 vehicles is assigned to discriminate 
between sparse and dense traffic. In other words, a vehicle continuously counts 
the surrounding vehicles in its communication range and calculates the average 
over time. If the average number of surrounding vehicles is lower than 30 then 
the traffic is considered sparse, otherwise it is considered dense. 

\subsection{CADS Evaluation}

\subsubsection{Location Privacy under GPA}
\begin{figure}
	\centering
	\includegraphics[scale=.6]{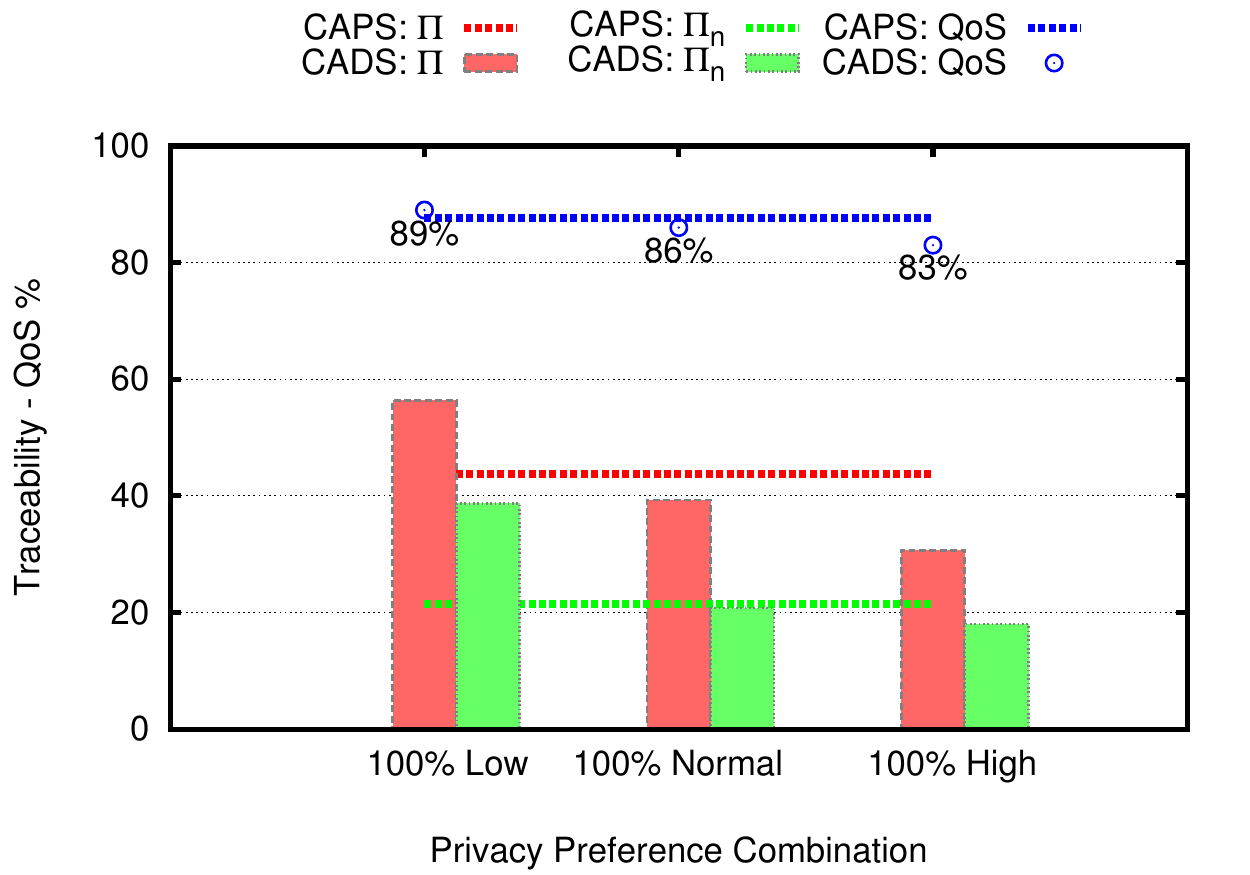} 
	\caption{The CADS evaluation when all vehicles use the same privacy 
	preference compared to the CAPS of 11 s max silent time.}
	\label{fig:cadsresall}
\end{figure}

CADS was evaluated under the GPA in two different scenarios. In the first 
scenario, all drivers select the same privacy preference whether low, normal or 
high level. Figure \ref{fig:cadsresall} displays the traceability, the 
normalized traceability and the quality of service of each privacy level. As a 
kind of comparison, the measurements for the CAPS scheme of 11 s maximum silent 
time are shown as dashed lines. The traceability and normalized traceability of 
CADS decrease when drivers select a higher privacy preference with a slight 
decrease in the QoS. Compared to CAPS, the CADS achieves a better compromise 
between traceability and QoS. Specifically, when a high privacy preference is 
used, the CADS achieves a 13\% lower traceability but with a slight decrease in 
QoS (only 4\%). When a low privacy preference is used, the QoS is enhanced by 
2\% while the normalized traceability is still lower than 40\%. In normal 
privacy preference, traceability is slightly decreased because of the 
adaptation of the parameters based on the traffic density.
These results confirm the validity and effectiveness of the 
context-adaptability to find a practical compromise between privacy preference 
and QoS. 

In the second scenario, we allow vehicles to select the preferred privacy level 
randomly based on given percentages. In this scenario, we aim to confirm that 
the privacy is more enhanced for vehicles that select a higher privacy level 
than the others. As the vehicles use a mix of privacy preferences, each privacy 
preference group is evaluated separately showing its traceability and 
normalized traceability. However, the QoS is evaluated over all vehicles, as 
lower-quality information obtained from vehicles that use a high privacy 
preference will affect other vehicles of lower privacy preferences and vice 
versa. In this scenario, we repeat each experiment five times with random 
selection of the privacy preference assigned to vehicles. 
%Correction ended here

\begin{figure}
	\centering
	\includegraphics[scale=.6]{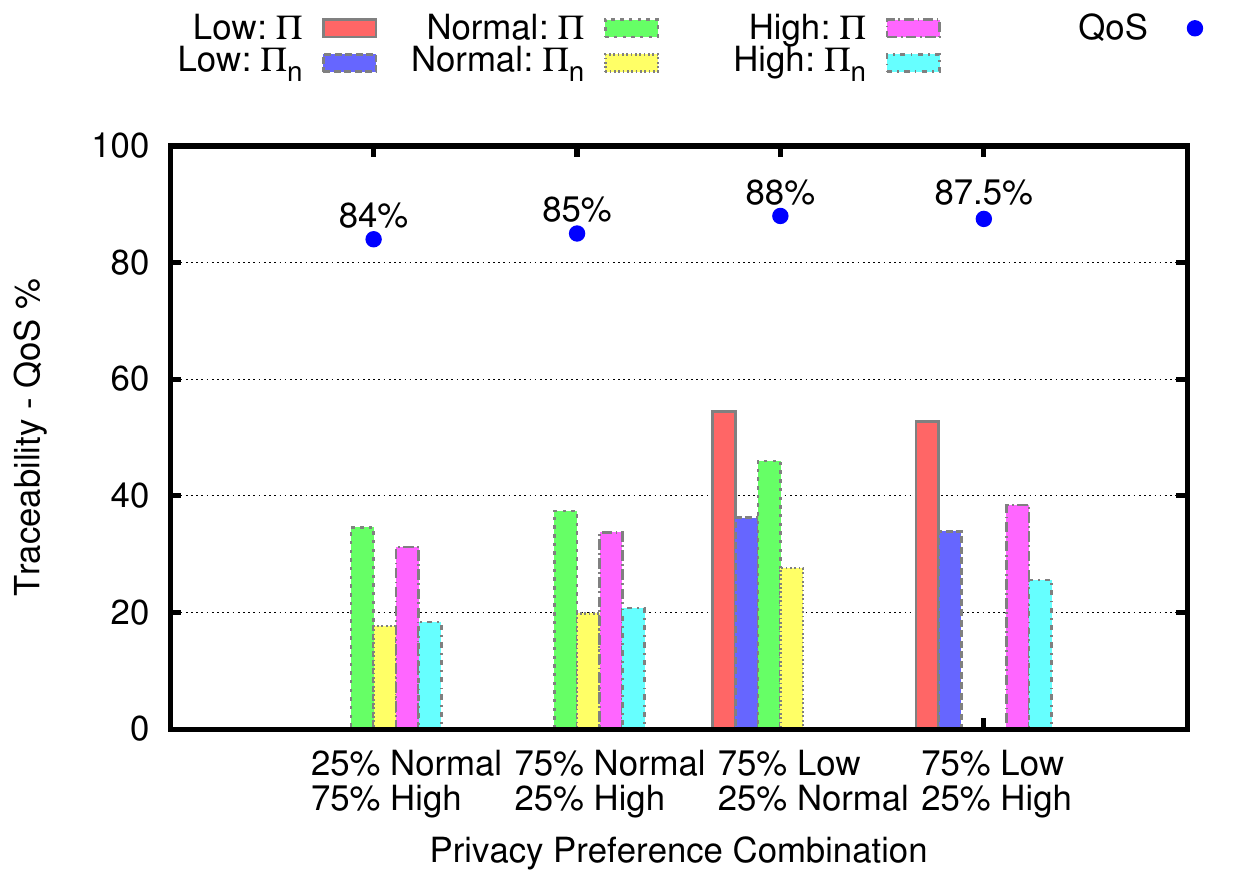} 
	\caption{The CADS evaluation when vehicles use a random privacy preference 
	based on the specified percentages.}
	\label{fig:cadsresmix}
\end{figure}
In the first and second experiments, 25\% and 75\% of vehicles use the normal 
privacy preference, respectively, while the rest uses the high privacy 
preference, as shown in Figure \ref{fig:cadsresmix}. Although both experiments 
employ swapped percentages of normal and high privacy levels, they achieve 
similar (normalized) traceability for both level groups with slight effect of 
the major group on the performance of the minor group. 

\begin{table}
	\centering
	\caption{CADS results under the LAA pseudonym depletion attack in sparse 
	sub-dataset (silent neighbor threshold = 1; 3967 vehicles)}
	\label{tab:laas}
	\begin{tabular}{r lllll}
		\toprule
		& \multicolumn{5}{c}{LAA strength} \\
		& No LAA & 1\%   & 3\%   & 5\%   & 10\%  \\ \cmidrule{1-6}
		Compromised vehicles             & 0      & 40    & 119   & 198   & 
		397   \\
		Concerned vehicles (victim)     & 2106   & 224   & 557   & 1041  & 
		1562  \\
		Avg. pseudonym lifetime (s)   & 114    & 88    & 85    & 80    & 
		74    \\
		Pseudonym change/Vehicle     & 1.3    & 1.8   & 1.8  & 1.8  & 1.9  
		\\ \cmidrule{1-6}
		$\Pi_n $ (\%) & 37    & 24 	  & 21   & 24     & 
		25 \\
		QoS (\%)                          & 88     & 87    & 85    & 83    & 
		79   \\ \bottomrule
	\end{tabular}
\end{table}
In the third and fourth experiments, 75\% of vehicles use the low privacy 
preference while the rest use normal and high levels, respectively. It is 
observable that the high level group in the fourth experiment achieves a lower 
traceability than that is achieved by the normal level group in the third 
experiment. Additionally, we notice that the high level group in the fourth 
experiment achieves slightly higher traceability than the same group in the 
second experiment. This result may attributed to the major privacy preference 
group being low-level in the fourth experiment but normal-level in the second. 
Regarding the QoS, we notice that it follows the QoS of the major group with a 
slight effect from the minor. For example, the QoS in the first experiment is 
higher 1\% than that in the ``100\% high-privacy'' experiment, and the QoS in 
the fourth experiment is lower 1.5\% than that in the ``100\% low-privacy'' 
experiment. From all these observations, we can conclude that the traceability 
is mainly affected by the configured privacy level with a slight effect from 
the surrounding traffic. However, this change in traceability is compensated in 
the QoS. 

\subsubsection{Location Privacy under LAA}
The local active adversary (LAA) performs a pseudonyms depletion attack which 
tries to force victim vehicles to change pseudonyms as soon as possible. It is 
important to evaluate context-based schemes under this attack because these 
schemes change pseudonyms based on conditions that are external from the 
vehicle. Therefore, an adversary may try to mimic these conditions to force 
vehicles change pseudonyms frequently and deplete their pseudonyms pool. We 
simulate this attack by letting a random number of compromised vehicles drive 
within the road network. These vehicles act as LAA by changing their pseudonyms 
every 5 s and keep silent for 3 s and so on. This behavior is challenging the 
practicality of this attack because if the compromised vehicles change their 
pseudonyms, they will suffer from self-depletion in short time when they use 
authenticated pseudonyms. If they use fake pseudonyms or do not change 
pseudonyms but switch to silence frequently, surrounding vehicles can detect 
this behavior and abandon the compromised vehicles from affecting their 
decisions. Regardless of the practicability issues, we assume here that the 
compromised vehicles own infinite number of authenticated pseudonyms and is 
able to change it freely.

%They may keep their pseudonyms unchanged to avoid self-depletion, although 
%this leads to be easily detected.  
In the worst case scenario, a victim vehicle will change its pseudonym every 
minimum pseudonym time, but the CADS can reduce the effect of this attack 
through its parameter: the silent neighbor threshold. When the silent neighbor 
threshold is set to be more than one, the scheme requires several silent 
neighboring vehicles to switch to silence. This condition hinders the LAA 
attack since it is unlikely to have several LAA vehicles neighboring the victim 
vehicle. Also, CADS can employ the pseudonym time increment parameter to 
increase the minimum pseudonym time when the pseudonym is changed with a likely 
tracker confusion. 
%The CADS is slightly modified such that the pseudonym time increment is set to 
%120 s in the parameter table for all privacy preferences and traffic 
%densities. 

The CADS is evaluated against the LAA of different strengths in terms of the 
number of the compromised vehicles. The protection against this attack is 
measured by the number of pseudonym changes and the pseudonym lifetime made by 
vehicles on average. When calculating this metric, we considered only vehicles 
that met a LAA vehicle within 50 m radius for at least 15 s and changed their 
pseudonyms during simulation at least once. We selected the first and the last 
5 min of the realistic traces and run simulation five times for each LAA 
strength with different compromised vehicles selected randomly. We selected 2 
sub-datasets to show the effect of LAA on both sparse and dense traffic. These 
short traces will not affect the generality of the obtained results because we 
consider the pseudonym changing behavior rather than a full reconstruction of 
long traces. We tested two thresholds of silent neighbors of 1 and 2 vehicles 
where all vehicles choose the normal privacy preference. 

\begin{table}
	\centering
	\caption{CADS results under the LAA pseudonym depletion attack in dense 
	sub-dataset (silent neighbor threshold = 2; 7390 vehicles)}
	\label{tab:laad}
	\begin{tabular}{r lllll}
		\toprule
		& \multicolumn{5}{c}{LAA strength} \\
		& No LAA & 1\%   & 3\%   & 5\%   & 10\%  \\ \cmidrule{1-6}
		Compromised vehicles             & 0      & 74    & 222   & 370   & 
		739   \\
		Concerned vehicles (victim)     & 3526   & 744   & 2015   & 2946  & 
		3855  \\
		Avg. pseudonym lifetime (s)   & 156    & 142    & 132    & 122    & 
		103    \\
		Pseudonym change/Vehicle     & 1.1    & 1.2   & 1.2  & 1.3  & 1.4  
		\\ \cmidrule{1-6}
		$\Pi_n $ (\%) & 38    & 30  & 31   & 30     & 
		27 \\
		QoS (\%)                          & 91     & 90    & 89    & 88    & 
		86  \\ \bottomrule
	\end{tabular}
\end{table}
Table \ref{tab:laas} shows the average metrics obtained using a silent neighbor 
threshold of one for the sparse sub-dataset. Four LAA strengths along with the 
case of no LAA are evaluated. The number of the compromised vehicles and the 
concerned vehicles, on which the given metrics are calculated, are listed in 
the first two rows of Table \ref{tab:laas}. The concerned vehicles are those 
changed their pseudonyms at least once and refer to the victim vehicles when 
LAA is present or all vehicles for the no LAA case. The next two rows show the 
average pseudonym lifetime and the number of pseudonyms changed per vehicle. It 
can be observed that the victim vehicles changed pseudonyms 1.38 times more 
than the case of no LAA. This small increase in pseudonym changes cannot result 
in pseudonym depletion unless the LAA vehicles continuously follow the victim 
vehicles. Furthermore, we show the traceability and QoS metrics for each case. 
Interestingly, the normalized traceability metric $\Pi_n$ is decreased when the 
LAA is present because the compromised vehicles force surrounding vehicles to 
change pseudonyms. The increased pseudonym changes result in a decrease in QoS 
depending on the LAA strength. We repeated this experiment with a silent 
neighbor threshold of 2 but we found that the traceability is significantly 
increased because it is rarely to find two silent neighbors in this sparse 
traffic. \\

Table \ref{tab:laad} shows the average metrics obtained using a silent neighbor 
threshold of 2 for the dense sub-dataset. We use here a threshold of 2 because 
the traffic is dense and it is common to meet with a compromised vehicle 
repeatedly. We observe that the victim vehicles changed pseudonyms 1.27 times 
more than the case of no LAA at maximum. The same behavior of decreased 
traceability and slight reduction in QoS is also observed.\\

From these observations, we conclude that a weak LAA of small percent of 
compromised vehicles (e.g., up to 3\%) does not add a significant risk of 
pseudonyms depletion specially when setting the silent neighbor threshold to 
more than one. Also, this attack may hinder the threat of the GPA attack with a 
small impact on the QoS of safety applications.

\subsection{CADS Efficiency}
CADS was implemented using MATLAB as a centralized program, which operates on 
samples located in the communication range of each vehicle separately. We 
exploit the parallel for loop feature in MATLAB to iterate on vehicles 
asynchronously at every time step. We run our experiments on an Intel QuadCore 
i7-4800MQ @ 2.70GHz Hyper-threaded CPU. We calculate the running time of the 
CADS to process samples received by a vehicle in a single time step and average 
over all vehicles and time steps. We found that the average running time is 5 
ms for realistic traces. Note that this running time is obtained using a single 
thread, as the CADS code is basically sequential. Thus, this running time is 
reproducible on single-thread single-core CPUs of the given speed. Therefore, 
we can conclude that the CADS is efficient when high-end CPUs are used because 
the most frequent beaconing rate is 100 ms and the vehicle will have plenty of 
time to do other tasks. However, if lower-end CPUs are used inside vehicles, 
then further code optimization should be investigated. The memory is not an 
issue, as the CADS uses only a few hundreds of kBs for the Kalman filter tracks 
of the nearby vehicles. 

\section{Conclusion}
\label{sec:con}
We discussed context-aware (CAPS) and context-adaptive (CADS) privacy schemes 
for vehicular networks. They utilize a context monitoring module to track 
surrounding neighbors and identify adequate situations to change pseudonym and 
determine the effective length of silence period. In CADS, a driver can choose 
the desired privacy level and the scheme can automatically identify the 
appropriate parameters that match this desired level based on the real-time 
traffic density. Based on the experimental results, CADS can reduce 
traceability than the CAPS does when normal or high privacy levels are selected 
with a slight reduction in the QoS. Also, the CADS can preserve lowest 
traceability for vehicles that select a high privacy level even when they drive 
within a majority of vehicles selected a lower privacy level. In future work, 
we will compare CADS with advanced privacy schemes such as mix-zones. Also, we 
will investigate allowing vehicles measure the safety level in real-time to 
stop silence in critical situations, for example. Lastly, we will consider 
deploying CADS on a test platform for VANET such as NEC LinkBird-MX to measure 
and optimize its practical efficiency. 

\bibliographystyle{IEEEtran}
\bibliography{Bib}
\end{document}